\documentclass[12pt]{article}
\usepackage[margin=1.0in]{geometry}
\usepackage{setspace}
\RequirePackage[OT1]{fontenc}
\RequirePackage{amsthm,amsmath}
\RequirePackage[numbers]{natbib}
\RequirePackage[colorlinks,citecolor=blue,urlcolor=blue]{hyperref}
\usepackage{blindtext}
\usepackage{amsmath}
\usepackage{amsthm,amssymb}
\usepackage{amstext,verbatim}
\usepackage{bm}
\usepackage{mathrsfs,dsfont}
\usepackage{graphicx}
\usepackage{array}
\usepackage{natbib}
\usepackage{epsfig}
\usepackage{sectsty}
\usepackage{algorithm}
\usepackage{color}
\usepackage{longtable}
\usepackage{pifont}
\usepackage{rotating}
\usepackage{adjustbox}
\usepackage{appendix}
\usepackage{verbatim}

\theoremstyle{plain}
\newtheorem{theorem}{Theorem}

\newtheorem{lemma}{Lemma}

\newtheorem{definition}{Definition}

\newtheorem{condition}{Condition}

\newcommand{\fdr}{\mathrm{FDR}}
\newcommand{\fdp}{\mathrm{FDP}}
\newcommand{\mean}{\mathbb{E}}
\newcommand{\indep}{\rotatebox[origin=c]{90}{$\models$}}
\newcommand\smallO{
	\mathchoice
	{{\scriptstyle\mathcal{O}}}
	{{\scriptstyle\mathcal{O}}}
	{{\scriptscriptstyle\mathcal{O}}}
	{\scalebox{.6}{$\scriptscriptstyle\mathcal{O}$}}
}

\title{Controlling the False Discovery Rate for Binary Feature Selection via Knockoff}
\author{Yuxiang Xie, Gary Chan}
\begin{document}
\maketitle
\doublespacing

\begin{abstract}
	Variable selection has been widely used in data analysis for the past decades, and it becomes increasingly important in the Big Data era as there are usually hundreds of variables available in a dataset. To enhance interpretability of a model, identifying potentially relevant features is often a step before fitting all the features into a regression model. A good variable selection method should effectively control the fraction of false discoveries and ensure large enough power of its selection set. In a lot of contemporary data applications, a great portion of features are coded as binary variables. Binary features are widespread in many fields, from online controlled experiments to genome science to physical statistics. Although there has recently been a handful of literature for provable false discovery rate (FDR) control in variable selection, most of the theoretical analyses were based on some strong dependency assumption or Gaussian assumption among features. In this paper we propose a variable selection method in regression framework for selecting binary features.   
	Under mild conditions, we show that FDR is controlled exactly under a target level in a finite sample if the underlying distribution of the binary features is known. We show in simulations that FDR control is still attained when feature distribution is estimated from data. We also provide theoretical results on the power of our variables selection method in a linear regression model or a logistic regression model. In the restricted settings where competitors exist, we show in simulations and real data application on a HIV antiretroviral therapy dataset that our method has higher power than the competitor. 
\end{abstract}

\maketitle

\section{Introduction}
Generalized linear models, including linear regression model and logistic regression model, are widely used in statistical analysis of real data. In a regression framework, variable selection is one of the most popular tools for analyzing high dimensional data, in which a great number of features are available for modeling, while only a few of them are thought to be significantly associated with the response of interest. To enhance interpretability and predictability, it is crucial to identify the subset of relevant features before running a regression model. Many variable selection procedures with good theoretical properties have been proposed for the past two decades. For example, \citet{Tibshirani96} proposed Lasso penalized linear regression model, which uses an $l_1$ penalty. \citet{Fan01} proposed SCAD, a non-convex penalty, for variable selection. \citet{Zou05} proposed regularization and variable selection via elastic net. An important question about variable selection is how many features should be selected in the model. As a data-driven approach, the cross-validation method is commonly used for deciding the number of features selected (\citet{Shao93, Zhang93, Yu14}). However, most cross-validation methods do not guarantee the control of false discovery rate (FDR) for the selected features. 

\citet{barber15} proposes `Knockoff' to conduct variable selection and control the false discovery rate simultaneously. The original Knockoff procedure, though elegant and salient, has a couple of limitations: it assumes that the underlying model is Gaussian linear with homoscedasiticity and does not work for high dimensional setting (i.e. more features than the sample size). 
\citet{Candes17} then extends the Knockoff idea to a model free procedure (Model-X Knockoff) which allows the underlying model to be any type and also allows for high dimensional set-up. Instead of knowing the relationship between the response variable and the features, Model-X Knockoff requires the knowledge of the distribution of features. Shifting the burden of knowledge from the true regression model to the distribution of features is reasonable, particularly in the case where features are from case-control studies.  
\citet{fan17} shows that when the features are generated from a Gaussian graphical model, under some mild assumptions the Model-X Knockoff not only controls the false discovery rate, but also has asymptotic power equal one. \citet{Weinstein18} further conducts a power and prediction analysis for Knockoff using lasso statistics, and their analyses mainly focus on the cases where the distribution of features is continuous. Some interesting applications of Knockoffs can be found in \citet{gao18}, \citet{xiao17} and \citet{xie18}. 

Although there has been a handful of `Knockoff' methods, most of the theoretical analyses focus only on the case where the distribution of features is continuous. Nonetheless, binary datasets are also widespread in many fields, from online controlled experiments to genome science to physical statistics. When the features in a model all take binary values, it is not reasonable to assume normality of their distribution. \citet{Sesia18} has developed algorithm to sample Knockoff variables with the assumption that the features can be described by a hidden Markov model, but there is still a lack of methodology for extending the exact construction and theoretical analysis of Knockoff to binary features setting. 

Ising graphical model is a standard model of a phase transition for ferromagnetism in statistical mechanics, and it is very popular in modeling the pairwise interactions between binary variables via Ising model. In addition, multivariate Bernoulli model (\citet{Dai13}) is an extension of Ising graphical model, which further allows modeling clique effects among the binary variables. Therefore, rather than Gaussian graphical model or other continuous graphical models, it is more natural to assume that the binary features are generated from an Ising graphical model or a multivariate Bernoulli model.

\textbf{Our contributions.} Since there is no tailored method of applying the Knockoff idea to binary features in existing literature, not to mention a thorough power analysis, in this paper we close this gap by developing a Knockoff procedure for features following Ising distribution or multivariate Bernoulli distribution. In particular, we
\begin{enumerate}
	\item develop an exact construction of Knockoffs for binary features that are generated from Ising or multivariate Bernoulli models,
	
	\item provide theoretical analyses on the FDR control and asymptotic power of our Knockoff selection set,
	
	\item propose a second-order approximation construction to speed up the Knockoff procedure,
	
	\item confirm the practical utility of the proposed method by comparing it to existing Knockoff procedures in simulations and real data application.
\end{enumerate}

\section{Preliminaries}
\subsection{Model-X Knockoff} \label{subsec_knockoff}
The Model-X Knockoff procedure is a FDR-control variable selection method in a framework with a response variable $Y$ and multiple features $X=\left(X_1, \ldots, X_p \right)$.
\begin{definition}{(\citet{Candes17})}
	$\tilde{X}=\left(\tilde{X}_1, \ldots, \tilde{X}_p \right)$ are Model-X knockoffs for the original features $X=\left(X_1, \ldots, X_p \right)$ if 
	\begin{itemize}
		\item $\tilde{X} \indep X | Y$,
		
		\item and for any subset $S \subset \{1, \ldots, p\}$,
		\begin{equation} \label{mxkf_swap}
		\left(X, \tilde{X}\right)_{\text{swap(S)}} \overset{d}{=} \left(X, \tilde{X}\right),
		\end{equation}
		where $\left(X, \tilde{X}\right)_{\text{swap(S)}} $ means swapping the $X_j$ and $\tilde{X}_j$ for all $j \in S$. 
		
	\end{itemize}  
\end{definition}

Note that the Model-X Knockoff does not assume knowledge of the conditional distribution of $Y|X$ or the relationship between $Y$ and $X$. Instead, it does assume the joint distribution of the features is known. The exchangeability condition \eqref{mxkf_swap} is the key of the Knockoff procedure and most of its variants, and the technical difficulty in constructing $\tilde{X}$ is to ensure this exchangeability condition \eqref{mxkf_swap} to hold. \citet{Candes17} provide an exact construction of $\tilde{X}$ in the case where the features are Gaussian distributed. They also propose a second-order approximation for constructing knockoffs in the case where the features are not Gaussian, however their theoretical result of FDR control does not hold exactly for the approximation construction.

After constructing knockoff $\tilde{X}$, under generalized linear model of $Y$ given $X$, \citet{Candes17} propose to first solve a lasso type regression problem on the augmented design matrix $\mathbf{X^*}=\left[\mathbf{X} \text{ } \mathbf{\tilde{X}}\right]$, and denote the solution by $\hat{\beta}(\lambda)$, where the tuning parameter $\lambda$ is selected by cross-validation. Then set $Z_j=\left|\hat{\beta}_j(\lambda)\right|$ and $\tilde{Z}_{j}=\left|\hat{\beta}_{j+p}(\lambda)\right|$. The Lasso Coefficient Difference (LCD) statistic is defined to be
\begin{equation} \label{lcd}
W_j=Z_j-\tilde{Z}_j=\left|\hat{\beta}_j(\lambda)\right|-\left|\hat{\beta}_{j+p}(\lambda)\right|.
\end{equation} 

Let $\hat{\mathcal{S}}$ be the variable selection set and $\mathcal{S}$ be the set of non-zero coefficients in the true model. The false discovery rate (FDR) is defined to be $\mean\left[\fdp\right]$ where $\fdp=\frac{\left|\hat{\mathcal{S}} \cap \mathcal{S}^C \right|}{\left|\hat{\mathcal{S}}\right|}$. For a given $q \in (0, 1)$, choose a positive threshold $T$ as
\begin{equation} \label{threshold}
T=\min\left\{t>0:\frac{\#\left\{j:W_j\leq -t\right\}+1}{\#\left\{j:W_j\ge t \right\}} \leq q \right\},
\end{equation} 
and the Knockoff selected set $\hat{\mathcal{S}}=\left\{j:W_j\ge T \right\}$ controls the FDR at the level of $q$. 

\subsection{Ising Model and Multivariate Bernoulli Distribution}
Consider an Ising graphical model with $p$ nodes denoted by $X_j$, $1\leq j \leq p$. We assume in the rest of this paper that each $X_j$ takes either $+1$ or $0$, though our analysis is also applicable to $X_j$'s taking $+1$ or $-1$. The joint distribution of $X_j$'s takes the form
\begin{align} \label{ising_pdf}
&P_{\Theta} (X_1=x_1, \ldots, X_p=x_p) \nonumber\\
&=\frac{1}{Z(\Theta)} \exp\{\sum_{j=1,\ldots,p}\Theta_{jj}x_j+ \sum_{(j, j') \in \mathcal{E}} \Theta_{jj'} x_j x_{j'}\},
\end{align} 
where $Z(\Theta)$ is a normalization term. Given an Ising parameter matrix $\Theta \in \mathbb{R}^{p \times p}$, we can define an undirected graph $G=\left(\mathcal{V}, \mathcal{E}\right)$, where $\mathcal{V}=\left\{1,\ldots,p \right\}$, and $\left(j,j'\right) \in \mathcal{E}$  if and only if $\Theta_{jj'} \neq 0$ for $1\le j, j' \le p$ and $j\neq j'$. 

The Ising model was first adopted in physics \citep{Ising1925}. Following the terminology in physics, the $p$ nodes are $p$ magnetic dipoles, and the Ising parameter $\Theta_{jj'}$ is the coupling coefficient that describes the physical interaction between dipoles $j$ and $j'$ under the external magnetic field. 

It is worth noting that Ising model is a special case of Multivariate Bernoulli model, which has been extensively studied in \citet{Dai13}. The joint distribution of $X_j$'s following a Multivariate Bernoulli distribution takes the form
\begin{align} \label{mvb_pdf}
& \text{\indent}P_{\mathbf{f}}(X_1=x_1, \ldots, X_p=x_p) \nonumber\\
&=\frac{1}{b(\mathbf{f})} \exp\left\{ \sum_{r=1}^{p}\left(\sum_{1\leq j_1 <\ldots<j_r\leq p} f^{j_1\ldots j_r} B^{j_1\ldots j_r}(x) \right)  \right\},
\end{align}
where $b(\mathbf{f})$ is a normalizing constant, $B^{j_1 \ldots, j_r}(x)=x_{j_1}\ldots x_{j_r}$, and $f^{j_1\ldots j_r}$ are the natural parameters that have a bijective mapping to the general parameters $P_{\mathbf{f}}(x_1, \ldots, x_p)$ (\citep{Dai13}). For convenience, we use $\pi_{x_1 \ldots x_p}$ to denote $P_{\mathbf{f}}(x_1, \ldots, x_p)$ in the rest of the paper.

\subsection{Two Transformations} \label{2trans}
Suppose that $\left(X, \tilde{X}\right)$ follows a multivariate Bernoulli distribution with the joint probability $\pi_{x_1 \ldots x_p \tilde{x}_1 \ldots \tilde{x}_p}$. There are two important transformations of $\pi$ used in the log-linear regression models and the multivariate logistic regression models (Chapter 6 in \citet{mccullagh89}). The log-linear approach is based on the transformation $\pi \to \gamma$ defined by
\begin{eqnarray} \label{log_linear}
\gamma^{X_1}=\log \frac{\pi_{1*\ldots*}}{\pi_{0*\ldots*}}, \indent  \ldots, \indent \gamma^{\tilde{X}_p}=\log \frac{\pi_{*\ldots*1}}{\pi_{*\ldots*0}} \nonumber \\
\gamma^{X_1X_2}=\log \frac{\pi_{11*\ldots*}\pi_{00*\ldots*}}{\pi_{01*\ldots*}\pi_{10*\ldots*}}, \indent  \ldots, \indent \gamma^{\tilde{X}_{p-1}\tilde{X}_p}=\log \frac{\pi_{*\ldots*11}\pi_{*\ldots*00}}{\pi_{*\ldots*01}\pi_{*\ldots*10}} \nonumber \\
\vdots \nonumber \\
\gamma^{X_1\ldots \tilde{X}_p}=\log \frac{\prod \pi \text{ with even number of zeros in subscript}}{\prod \pi \text{ with odd number of zeros in subscript}}, \nonumber
\end{eqnarray}
where $*$ denotes the geometric mean taken over the subscript. $\gamma$'s are related to conditional odds ratios. 

The multivariate logistic approach is based on the transformation $\pi \to \eta$ defined by
\begin{eqnarray} \label{multi_logistic}
\eta^{X_1}=\log \frac{\pi_{1+\ldots+}}{\pi_{0+\ldots+}}, \indent  \ldots, \indent \eta^{\tilde{X}_p}=\log \frac{\pi_{+\ldots+1}}{\pi_{+\ldots+0}} \nonumber \\
\eta^{X_1X_2}=\log \frac{\pi_{11+\ldots+}\pi_{00+\ldots+}}{\pi_{01+\ldots+}\pi_{10+\ldots+}}, \indent  \ldots, \indent \eta^{\tilde{X}_{p-1}\tilde{X}_p}=\log \frac{\pi_{+\ldots+11}\pi_{+\ldots+00}}{\pi_{+\ldots+01}\pi_{+\ldots+10}} \nonumber \\
\vdots \nonumber \\
\eta^{X_1\ldots \tilde{X}_p}=\log \frac{\prod \pi \text{ with even number of zeros in subscript}}{\prod \pi \text{ with odd number of zeros in subscript}}, \nonumber
\end{eqnarray}
where $+$ denotes the summation over the subscript. $\eta$'s are related to lower dimensional marginal probabilities.

\citet{glonek96} has studied the mapping $\pi \to \left(\eta, \gamma\right)$, where $\left(\eta, \gamma\right)$ is a mixed parametrization.
The combination of $\eta$ and $\gamma$ needs to follow the hierarchy principle in \citep{glonek96}. The mapping $\pi \to \left(\eta, \gamma\right)$ is invertible under mild conditions and \citet{glonek96} has proposed an inversion algorithm for getting $\pi$ from $\left(\eta, \gamma\right)$.

\section{Binary Knockoff Procedure}
In this section, we propose a method for binary feature selection, which can control FDR at a pre-specified level and maintain large enough power at the same time. 

Given $n$ random draws of $X=(X_{1},\ldots, X_{p})$ from a binary feature distribution $F_X$ and $n$ random draws of $Y$ from a response distribution $F_Y$, we want to select features from $(X_1, \ldots, X_p)$ that are significantly associated with response $Y$, while keeping FDR below a target level. We assume the feature distribution of $X$ is known, but we assume neither knowledge of the distribution of $Y$ nor knowledge of the relationship between $Y$ and $X$.

We assume in the rest of the paper that the features are generated from an Ising model, though the construction of Knockoffs and the theoretical results in this section are also applicable to multivariate Bernoulli features. We focus on the case of Ising features because it has more practical applications due to the fact that there exists many estimation methods for parameters in Ising models. 

The main contribution of our proposal is the construction of binary knockoffs. After constructing knockoff $\tilde{X}$ for the original $X$, we follow the same procedure of Model-X Knockoff in Section 2.1 to obtain the knockoff selection set $\hat{\mathcal{S}}$.

\subsection{Exact Construction of Binary Knockoffs}
Suppose that we have $n$ independent draws from $X=\left(X_1, \ldots, X_p\right)$ following an Ising model with known coupling coefficient parameter $\Theta^*$. Let $\mathbf{X} \in \mathbb{R}^{n \times p}$ be the design matrix such that each row is a draw. We first present an exact construction of Binary Knockoffs that satisfies the exchangeability condition \eqref{mxkf_swap}, which is the key to FDR control of Knockoff procedure. Our proposed exact construction of Binary Knockoffs takes the following steps:

\begin{itemize}
	\item \textbf{Step 1:} Choose a mixed parametrization $\left(\eta, \gamma\right)$ for $\pi_{x_1 \ldots x_p \tilde{x}_1 \ldots \tilde{x}_p}$.
	
	\item \textbf{Step 2:} Calculate a part of $\eta$ using the given Ising parameters of $X$.
	
	\item \textbf{Step 3:} Assign values to the rest of $\eta$ and $\gamma$ to ensure exchangeability condition \eqref{mxkf_swap} of $\left(X, \tilde{X}\right)$.
	
	\item \textbf{Step 4:} Invert the mapping $\pi \to (\eta, \gamma)$ to get $\pi$ from constructed $(\eta, \gamma)$. 
	
	\item \textbf{Step 5:} Obtain the conditional distribution $\tilde{X} | X$ from $\pi$ (joint distribution) and the known distribution of $X$ (marginal distribution). 
	
	\item \textbf{Step 6:} Sample knockoffs by using the conditional distribution $\tilde{X} | X$.
\end{itemize}

The inversion algorithm used in Step 4 can be found in \citet{glonek96}. Steps 5--6 are simple in theory. Our main effort is put on Steps 1--3 as we need the exchangeability condition to hold for FDR control. 

\textbf{Step 1.} Note that $\left(X_1, \ldots, X_p, \tilde{X}_1, \ldots \tilde{X}_p\right)$ contains $2p$ variables including knockoffs. Let the index set $\{1, \ldots 2p\}$ correspond to the order of $\left(X, \tilde{X}\right)$. To better illustrate the choice of a mixed parametrization, we use $\xi$ to denote a combination of $(\eta, \gamma)$. 

For any index subset $I \subset \{1, \ldots, 2p\}$ where $|I|\leq p$, we choose $\xi^{\left(X, \tilde{X}\right)_{I}}=\eta^{\left(X, \tilde{X}\right)_{I}}$, where $\left(X, \tilde{X}\right)_{I}$ corresponds to the superscript of $\eta$ defined in Section~\ref{2trans}.

For any index subset $J \subset \{1, \ldots, 2p\}$ where $|J|> p$, we choose $\xi^{\left(X, \tilde{X}\right)_{J}}=\gamma^{\left(X, \tilde{X}\right)_{J}}$, where $\left(X, \tilde{X}\right)_{J}$ corresponds to the superscript of $\gamma$ defined in Section~\ref{2trans}.

For example, when $p=2$, we choose a mapping from $\pi$ to a mixed $(\eta, \gamma)$ as 
\begin{align} \label{example_p2}
\pi \to (\eta^{X_1}, \eta^{X_2}, \eta^{\tilde{X}_1}, \eta^{\tilde{X}_2}, \eta^{X_1X_2}, \eta^{X_1\tilde{X}_1}, \eta^{X_1\tilde{X}_2}, \eta^{X_2\tilde{X}_1}, \eta^{X_2\tilde{X}_2}, \eta^{\tilde{X}_1\tilde{X}_2},  \nonumber \\
\gamma^{X_1X_2\tilde{X}_1}, \gamma^{X_1X_2\tilde{X}_2}, \gamma^{X_1\tilde{X}_1\tilde{X}_2}, \gamma^{X_2\tilde{X}_1\tilde{X}_2}, \gamma^{X_1X_2\tilde{X}_1\tilde{X}_2}). 
\end{align}  

This type of combination of $(\eta, \gamma)$ satisfies the hierarchy principle in \citep{glonek96}, thus the inversion algorithm in \citep{glonek96} is applicable to it.

\textbf{Step 2.} Multivariate Bernoulli model is an extension of Ising model with $f^{jj'}=\Theta_{jj'}$ and $f^{\mathbf{J}}=0$ for $|\mathbf{J}|>2$, where $f^{\mathbf{J}}$ are the natural parameters in \eqref{mvb_pdf} and $\Theta_{jj'}$ are the Ising parameters in \eqref{ising_pdf}. In addition, there is a bijective mapping between the natural parameters $f$ and the joint probabilities (i.e. general parameters) $\pi$ of a multivariate Bernoulli model. Therefore, given the Ising parameters of $X$, we are able to calculate $\pi_{\mathcal{I}}$ for any subset $\mathcal{I}$ of the power set of $\left\{x_1, \ldots, x_p\right\}$. The bijective transformation formula is explicitly stated in \citet{Dai13}.

Continue using the example in \eqref{example_p2}. We are able to calculate $\eta^{X_1}, \eta^{X_2},$ and $\eta^{X_1X_2}$, because $\eta^{X_1}, \eta^{X_2}, \eta^{X_1X_2}$ relate to the lower dimensional marginal probabilities $\pi_{x_1}, \pi_{x_2}, \pi_{x_1x_2}$, which can be calculated from the given Ising parameters of $X$. We will use these $\eta$ values in Step 3.

\textbf{Step 3.} The objective of our construction is to ensure exchangeability condition \eqref{mxkf_swap} of $\left(X, \tilde{X}\right)$. It requires appropriate assignment of $\eta$ and $\gamma$ values that are used for inverting back to $\pi$. 

First, consider the $\eta$ part. For any index subset $\mathcal{I} \subset \{1, \ldots, 2p\}$ where $|\mathcal{I}|\leq p$ and for any swapping index subset $S \subset \{1, \ldots, p\}$, we set 
\begin{equation} \label{exchangeable}
\eta^{\left(X, \tilde{X}\right)_{\mathcal{I}}}=\eta^{\left\{\left(X, \tilde{X}\right)_{\text{swap(S)}}\right\}_{\mathcal{I}}}
\end{equation}
by using the calculated $\eta$ values in Step 2. In the example \eqref{example_p2}, satisfying the condition \eqref{exchangeable} is equivalent to setting $\eta^{\tilde{X}_1}=\eta^{X_1}, \eta^{\tilde{X}_2}=\eta^{X_2}$, and  $\eta^{X_1X_2}=\eta^{\tilde{X}_1\tilde{X}_2}=\eta^{X_1\tilde{X}_2}=\eta^{\tilde{X}_1X_2}$.
Note that some $\eta$ values are non-identifiable, for example, $\eta^{X_1\tilde{X}_1}, \eta^{X_2\tilde{X}_2}$ in \eqref{example_p2}. We can simply set them to be some arbitrary values like zeros as long as \eqref{exchangeable} holds.

Next, consider the $\gamma$ part. We propose to set all $\gamma$ to be some constant $C$. We recommend trying $C=0$ when running the inversion algorithm in \citep{glonek96}, since it slightly simplify one step of the algorithm.

Using the example in $\eqref{example_p2}$ one more time, we may consider an inverse mapping of 
\begin{align} \label{example_p2_0}
\pi \to &(\eta^{X_1}, \eta^{X_2}, \eta^{X_1}, \eta^{X_2}, \nonumber \\
& \eta^{X_1X_2}, 0, \eta^{X_1X_2}, \eta^{X_1X_2}, 0, \eta^{X_1X_2},  \nonumber \\
& 0, 0, 0, 0, 0),
\end{align}  
and use the inversion algorithm in \citet{glonek96} to get $\pi_{x_1 x_2 \tilde{x}_1 \tilde{x}_2}$.

It is easy to check that our assignment of $\eta$ and $\gamma$ leads to a set of joint probabilities $\pi$ satisfying the exchangeability condition \eqref{mxkf_swap} of $\left(X, \tilde{X}\right)$. Furthermore, the construction of $\tilde{X}$ does not involve the response variable $Y$. Therefore, the Binary Knockoffs $\tilde{X}$ generated from our exact construction satisfy the two conditions for Model-X knockoffs, and consequently inherit the desirable properties of Model-X knockoffs. This leads to the following theoretical results.

\subsection{Theoretical Results}
After constructing knockoffs  $\tilde{X}$, we calculate LCD statistic $W_j$ and threshold $T$ (depends on a pre-specified FDR control level $q$) following the same manner in \citet{Candes17}.  $\hat{\mathcal{S}}=\left\{j:W_j\ge T \right\}$ is the Knockoff selection set. We first present the result for FDR control.

\begin{theorem}\label{thm1}
	Given the Ising features $X$ with known parameters $\Theta^*$, and using the exact construction for sampling knockoffs $\tilde{X}$, the Knockoff selected set $\hat{\mathcal{S}}=\left\{j:W_j\ge T \right\}$ controls the FDR at a pre-specified level $q$. 
	In addition, this FDR-control result is non-asymptotic and holds without knowledge of the underlying relationship between the response $Y$ and the features $X$.
\end{theorem}
\begin{proof}
	Since the Binary Knockoffs $\tilde{X}$ satisfy the Model-X knockoffs conditions, by Lemma 2 and Lemma 3 in \citet{Candes17}, the signs of the null statistics $\left\{W_j: \beta_j=0\right\}$ for $j=1,\ldots,p$ are distributed as random coin flips. Hence, following the same arguments in the proof of Theorems 1 and 2 in \citet{barber15}, our Knockoff procedure controls the false discovery rate at a pre-specified level.
\end{proof}

The advantages of the Knockoff procedure are obvious based on Theorem~\ref{thm1}: the FDR control result holds in finite samples and it works even if the model is mis-specified. 

In addition to the FDR control, we also provide analyses on asymptotic power of the Binary Knockoff procedure. In contrast to the FDR control analysis, the power analysis requires knowledge of the true model. We first consider the case where the true relationship between $Y$ and $X$ is linear: 
\begin{equation}
Y=X\beta+\epsilon, \nonumber
\end{equation}
where $\beta$ is the unknown true coefficient vector and $\epsilon$ is an error term.

Denote $X^*$ to be $\left(X, \tilde{X}\right)$, and $\mathbf{X}^*=[\mathbf{X} \text{ } \mathbf{\tilde{X}}] \in \mathbb{R}^{n \times 2p}$ to be the augmented design matrix. Let $|\mathcal{S}|=s$, where $\mathcal{S}$ is the set of non-zero coefficients in the true model. Let $q$ be the pre-specified level of FDR that we want to control via Knockoff. Let $\hat{\beta} \in \mathbb{R}^{2p}$ be the augmented coefficient estimates from running a Lasso regression using $Y$ and $\mathbf{X^*}$. Note that the augmented true coefficients $\beta_T$ is equal to $[\beta^T, \mathbf{0}^T]^T$, because $\tilde{X}$ is constructed without looking at $Y$, thus irrelevant to Y.

To facilitate the power analysis, we impose the following regularity assumptions:
\begin{itemize}
	\item \textbf{Condition 1:} The error components of $\epsilon$ are $i.i.d$ with a sub-Gaussian distribution.
	
	\item \textbf{Condition 2:} As $n$ increases, it holds that $\left(\frac{n}{\log p}\right)^\frac{1}{2}\min_{j \in \mathcal{S}}|\beta_{j}| \xrightarrow{} \infty$.
	
	\item \textbf{Condition 3:} With asymptotic probability one, $|\hat{\mathcal{S}}|\ge cs$ for some constant $c\in \left(2(qs)^{-1}, 1\right)$. 
	
	\item \textbf{Condition 4:} Let $\Sigma_0=\mean\left[X^{*T}X^*\right]$ and $\Sigma_0$ satisfies compatibility condition with some constant $\phi_{\Sigma_0}>0$, i.e.
	\begin{equation} \label{compability_cond}
	||\alpha_S||_1^2 \leq \frac{s\alpha^T \Sigma_0 \alpha}{\phi_{\Sigma_0}^2} 
	\end{equation}
	for all vectors $\alpha$ satisfying $||\alpha_{\mathcal{S}^C}||_1 \leq 3 ||\alpha_\mathcal{S}||_1$, where $\mathcal{S}^C$ is the complement set of $\mathcal{S}$. 
\end{itemize}

The error term $\epsilon$ does not need to follow exactly a sub-Gaussian distribution. We need a concentration inequality of sub-Gaussian distribution in the proof. Any other distributions with similar concentration inequalities can replace the sub-Gaussian in Condition 1. Condition 2 ensures the asymptotic power of Lasso to be one. This condition is needed since the Knockoff procedure uses Lasso in variable selection, so its asymptotic power is upper bounded by Lasso. Condition 3 puts a lower bound on the number of selected features. The Conditions 1--3 are exactly same as the ones in the asymptotic power analysis of Model-X Knockoff in \citet{fan17}. The analysis in \citet{fan17} is based on the assumption of Gaussian features. In contrast, $X^*$ is binary in our case and follows a multivariate Bernoulli distribution by construction. In order to obtain error bounds of Lasso results without Gaussian assumption, we further assume Condition 4 that imposes constraints on the smallest eigenvalue of the covariance $\Sigma_0$. It is reasonable to assume such condition in power analysis, as many theories on Lasso require similar restriction on the smallest eigenvalue of covariance.

\begin{theorem}\label{thm_pwr_linear}
	Assume that Condition 1--4 hold. Use the exact construction to obtain Binary Knockoffs $\tilde{X}$ and follow the Model-X Knockoff procedure to get Knockoff selection set $\hat{\mathcal{S}}$. With asymptotic probability one, $\frac{\left|\hat{\mathcal{S}} \cap \mathcal{S} \right|}{\left|\mathcal{S}\right|} \ge 1-\mathit{O}(a_n^{-1})$ for some $a_n \rightarrow \infty$, i.e. Power($\hat{\mathcal{S}}$) $\rightarrow 1$ as $n \rightarrow \infty$.
\end{theorem}

In order to prove Theorem~\ref{thm_pwr_linear}, we need the following Lemma~\ref{lemma_condition1} and Lemma~\ref{lemma_condition4}.
\begin{lemma} \label{lemma_condition1}
	Assume that $\mathbf{X^*} \in \mathbb{R}^{n \times 2p}$ has independent rows with all values being 0 or 1, and $\epsilon=(\epsilon_1, \ldots, \epsilon_n)$ are i.i.d sub-Gaussian components. 
	Then we have 
	\begin{equation} \label{eq_lemma_condition1}
	Pr\left(\left|\left|\frac{1}{n}\left(\mathbf{X^*}\right)^T\epsilon\right|\right|_{\infty}\leq C_2 \sqrt{(\log p)/n}\right) \ge 1-p^{-C_3}
	\end{equation}
	for large enough constant $C_2>0$ and some constant $C_3>0$.
\end{lemma}
\begin{proof}{(Lemma~\ref{lemma_condition1})}
	Since $X^*_{ij}=0$ or $1$ and $\epsilon=(\epsilon_1, \ldots, \epsilon_n)$ are i.i.d sub-Gaussian components by assumption, for $t>0$ we have
	\begin{equation}
	Pr\left(\left|\epsilon_i X^*_{ij}\right|>t\right) \leq Pr\left(\left|\epsilon_i\right|>t\right)\leq C_1\exp\left(-C_1^{-1}t^2\right).
	\end{equation}
	Thus by Lemma 6 in \citet{fan16}, we have 
	\begin{equation}
	Pr\left(\left|\frac{1}{n}\sum_{i=1}^{n}\epsilon_i X^*_{ij}\right|>v\right)\leq \tilde{C}_1 \exp\left(-\tilde{C}_1nv^2\right)
	\end{equation}
	for some $\tilde{C}_1>0$ and all $0<v<1$. Hence
	\begin{align}
	1-Pr\left(\left|\left|\frac{1}{n}\left(\mathbf{X^*}\right)^T\epsilon\right|\right|_{\infty}\leq v\right)&=Pr\left(\left|\left|\frac{1}{n}\left(\mathbf{X^*}\right)^T\epsilon\right|\right|_{\infty}> v\right) \\
	&=Pr\left(\max_{1\leq j\leq 2p}\left|\frac{1}{n} \epsilon^T \mathbf{X^*}_j \right|> v\right) \\
	&\leq 2p\tilde{C}_1 \exp\left(-\tilde{C}_1nv^2\right).
	\end{align}
	Substituting $v=C\sqrt{(\log p)/n}$ into the above inequality and taking large enough $C_2$, we have the stated result in Lemma~\ref{lemma_condition1}. 
\end{proof}

Note that our knockoff construction method ensures that $X^{*}$ is binary. In addition, Condition 1 implies the inequality of the error term are sub-Gaussian as assumed in Lemma~\ref{lemma_condition1}. Therefore, Lemma~\ref{lemma_condition1} always holds in our setup. 

Based on the result of Lemma~\ref{lemma_condition1}, combining with the basic inequality of Lasso regression, we can derive $\left|\left|\left(\hat{\beta}(\lambda)-\beta_T\right)_{\mathcal{S}^C}\right|\right|_1 \leq 3\left|\left|\left(\hat{\beta}(\lambda)-\beta_T\right)_{\mathcal{S}}\right|\right|_1$ with high probability.

\begin{lemma} \label{lemma_condition4}
	With high probability, Condition 4 implies the compatibility condition for $\Sigma_1=X^{*T}X^*$ with some constant $\phi_{\Sigma_1}>0$
\end{lemma}

\begin{proof}{(Lemma~\ref{lemma_condition4})}
	Note that $Z:=\Sigma_1-\Sigma_0=\Sigma_1-\mean\left[\Sigma_1\right]$, so $Z_{jk}=\frac{1}{n}\left(\sum_{i=1}^{n}Z_{jk}^{(i)}\right)$ where each $Z_{jk}^{(i)}$ is zero-mean and bounded (since $|Z_{jk}^{(i)}|\leq 2$). By the Azuma-Hoeffding bound, 
	\begin{equation}
	P\left((Z_{jk})^2\ge \lambda^2\right)=P\left(\left|\frac{1}{n}\left(\sum_{i=1}^{n}Z_{jk}^{(i)}\right)\right|\ge \lambda \right) \leq 2\exp\left(-\frac{\lambda^2 n}{32}\right).
	\end{equation}
	Therefore, $||\Sigma_1-\Sigma_0||_\infty \leq \lambda$ holds with high probability.
	
	Given $||\Sigma_1-\Sigma_0||_\infty \leq \lambda$, by \citet{bv11} Lemma 6.17, for all $\alpha$ s.t. $||\alpha_{\mathcal{S}^C}||_1 \leq 3 ||\alpha_\mathcal{S}||_1$ and $\Sigma_0$-compatibility condition holds, we have 
	\begin{equation}
	\left|\frac{\alpha^T\Sigma_1\alpha}{\alpha^T \Sigma_0 \alpha}-1\right|\leq \frac{16\lambda s}{\phi_{\Sigma_0}^2}.
	\end{equation}
	By \citet{bv11} Corollary 6.8, 
	then $\Sigma_1$-compatibility condition holds with $\phi_{\Sigma_1}^2\ge \phi_{\Sigma_0}^2/2$.
\end{proof}

Since $\Sigma_1$-compatibility condition is implied by Condition 4 with high probability, then by Theorem 6.1 in \citet{bv11} and the result from Lemma~\ref{lemma_condition1} that $\left|\left|\left(\hat{\beta}(\lambda)-\beta_T\right)_{S^c}\right|\right|_1 \leq 3\left|\left|\left(\hat{\beta}(\lambda)-\beta_T\right)_{S}\right|\right|_1$ with high probability, we have $\left|\left|\hat{\beta}(\lambda)-\beta_T\right|\right|_1=\mathcal{O}(s\lambda) $ where $\lambda=C\sqrt{(\log p)/n}$ with high probability, for some constant $C>0$, which will be used in the proof of Theorem~\ref{thm_pwr_linear}. 

\begin{proof}{(Theorem~\ref{thm_pwr_linear})}
	Now we start proving the main result in Theorem~\ref{thm_pwr_linear} by mimicking the way of proof of Theorem 3 in \citet{fan17}. Denote $W_j$ to be the LCD based on $\hat{\beta}(\lambda)$, and let $\left|W_{(1)}\right| \ge \ldots \ge \left|W_{(p)}\right|$ be the ordered knockoff statistics according to absolute size. Denote $j^*$ the index such that $\left|W_{j^*}\right|=T$, where $T$ is the threshold defined in \eqref{threshold}. It holds that $-T<\left|W_{j^*+1}\right|\leq 0$.
	
	\textbf{Case 1:} For the case of $W_{(j^*+1)}=0$, we have $W_{k}=0$ for $k=j^*+1, \ldots, p$. Then the index set $\left\{j: W_j\neq 0\right\}$ is same as the index set of $\hat{\mathcal{S}}$ selected by the Knockoff procedure. 
	We have
	\begin{equation} \label{case1_1}
	\left\{1, \ldots, p \right\} \backslash S_1 \subset \hat{\mathcal{S}},
	\end{equation}
	where $S_1=\left\{1 \leq j \leq p: \hat{\beta}_j(\lambda)=0\right\}$.
	
	We have shown in Lemma~\ref{lemma_condition1} and Lemma~\ref{lemma_condition4} that with high probability $\left|\left|\hat{\beta}(\lambda)-\beta_T\right|\right|_1=\mathcal{O}(s\lambda) $. Then we have
	\begin{align} \label{case1_2}
	\mathcal{O}(s\lambda) &=  \left|\left|\hat{\beta}(\lambda)-\beta_T\right|\right|_1 \ge \sum_{j\in S_1 \cap \mathcal{S}} \left|\hat{\beta}_j(\lambda)-\beta_{T,j}\right|=\sum_{j\in S_1 \cap \mathcal{S}} \left|\beta_{T,j}\right| \nonumber \\
	&\ge\left|S_1 \cap \mathcal{S}\right|\min_{j\in \mathcal{S}}\left|\beta_{T,j}\right|.
	\end{align}
	
	Since $\beta_{0,j}=\beta_{T,j}$ for $1\leq j\leq p$, by Condition 2 and $\lambda=\mathcal{O}(\sqrt{\frac{\log p}{n}})$, we can derive from \eqref{case1_2} that $\left|S_1 \cap \mathcal{S}\right|=\smallO(s)$, where $s=\left|\mathcal{S}\right|$. Also note that $\left|\left(\left\{1, \ldots, p \right\} \backslash S_1\right) \cap \mathcal{S} \right| \ge \left|\mathcal{S}\right|-\left|S_1-\mathcal{S}\right|=\left(1-\smallO(1)\right)s$. Together with \eqref{case1_1}, we obtain 
	\begin{equation} \label{case1_3}
	\left|\hat{\mathcal{S}} \cap \mathcal{S}\right| \ge \left(1-\smallO(1)\right)s.
	\end{equation}
	Therefore, with asymptotic probability one, we have $\frac{\left|\hat{\mathcal{S}} \cap \mathcal{S}\right|}{s} \ge 1-\smallO(1)$.

	\textbf{Case 2:} For the case of $-T< \left|W_{j^*+1}\right| < 0$, we first note that
	\begin{equation} \label{case2_1}
	\frac{\left|\left\{j:W_j \leq -T\right\}\right|+2}{\left|\left\{j:W_j\ge T \right\}\right|} > q,
	\end{equation}
	where $q$ is the pre-specified FDR control level. Then by Condition 3 together with $\eqref{case2_1}$, we have $\left|\left\{j:W_j \leq -T\right\}\right| > q \left|\left\{j:W_j\ge T \right\}\right| -2 \ge qcs-2$ with asymptotic probability one. In addition, in this case, $\left|\hat{\beta}_{j+p}(\lambda)\right| \ge T$ for all $j$ such that $W_j \leq -T$. Again, using the result of $\left|\left|\hat{\beta}(\lambda)-\beta_T\right|\right|_1$ from Lemma~\ref{lemma_condition1}--\ref{lemma_condition4}, we obtain
	\begin{equation} \label{case2_2}
	\mathcal{O}(s\lambda) =  \left|\left|\hat{\beta}(\lambda)-\beta_T\right|\right|_1 \ge \sum_{j:W_j \leq -T} \left|\hat{\beta}_{j+p}(\lambda)\right| \ge T\left|\left\{j:W_j \leq -T \right\}\right|.
	\end{equation}
	Therefore, $\mathcal{O}(s\lambda) \ge T(qcs-2)$, thus $T \leq \mathcal{O}(\lambda)$.
	
	On the other hand, we have
	\begin{align} \label{case2_3}
	\mathcal{O}(s\lambda) &=  \left|\left|\hat{\beta}(\lambda)-\beta_T\right|\right|_1 = \sum_{j=1}^{p} \left(\left|\hat{\beta}_j(\lambda)-\beta_{T,j}\right|+\left|\hat{\beta}_{j+p}(\lambda)\right|\right) \nonumber \\
	&\ge \sum_{j\in \mathcal{S} \cap \left(\hat{\mathcal{S}}\right)^C} \left(\left|\hat{\beta}_j(\lambda)-\beta_{T,j}\right|+\left|\hat{\beta}_{j}(\lambda)\right|-T\right).
	\end{align}
	
	By Condition 2, we have $\min_{j \in \mathcal{S}}\left|\beta_{0, j}\right| \ge \tau_n \lambda$ for some $\tau_n \rightarrow \infty$. Therefore, by \eqref{case2_3} and triangle inequality, we get 
	\begin{equation}
	\mathcal{O}(s\lambda) \ge \sum_{j\in \mathcal{S} \cap \left(\hat{\mathcal{S}}\right)^C} \left(\left|\beta_{0, j}-T\right| \right) \ge (\lambda \tau_n -T)\left|\left\{j \in \mathcal{S} \cap \left(\hat{\mathcal{S}}\right)^C \right\}\right|.
	\end{equation}
	
	With some algebra, we can conclude that $\frac{\left|\hat{\mathcal{S}} \cap \mathcal{S}\right|}{s} \ge 1-\smallO(1)$.
	
	By combining the results in two cases, we complete the proof of Theorem~\ref{thm_pwr_linear}.
\end{proof}

Theorem~\ref{thm_pwr_linear} shows that under some mild conditions the asymptotic power of our knockoff procedure approaches to one when the underlying model is a linear regression model. We also provide power analysis in a logistic regression model setup. Suppose that $Y\in \left\{-1, 1\right\}$ and $X$ are generated from an Ising model with parameters $\Theta^*$. Consider the true model to be 
\begin{equation}
P_{\beta_T}(Y|X^*)=\frac{\exp\left\{Y X^*\beta_T \right\}}{\exp\left\{Y X^*\beta_T \right\}+1}. \nonumber
\end{equation} 

Let $Q^*=\mean_{\beta_T} \left\{\triangledown^2 \log P_{\beta_T}\left[Y|X^* \right] \right\}$, the Fisher information matrix associated with the conditional probability distribution of $Y|X^*$. Let $Q_{\mathcal{S}\mathcal{S}}^*$ be the sub-matrix of $Q^*$ indexed by the true non-zero coefficient set $\mathcal{S}$.

To facilitate the analysis, we further impose the following basic regularity assumptions:

\begin{itemize}
	\item \textbf{Condition 5:} There exists some constant $C_{\min}>0$ s.t. the minimum eigenvalue of $Q_{\mathcal{S}\mathcal{S}}^*\ge C_{\min}$ and the maximum eigenvalue of $\mean\left[X^{*T}X^*\right] \leq D_{\max}$ for some positive constant $D_{\max}$.
	
	\item \textbf{Condition 6:} $\left|\left| |Q_{\mathcal{S}^C \mathcal{S}}^*(Q_{\mathcal{S}\mathcal{S}}^*)^{-1}|  \right|\right|_{\infty} \leq 1-\alpha$ for some $\alpha \in (0, 1]$. 
\end{itemize}

The first part of Condition 5 puts a lower bound on the eigenvalues of the Fisher information matrix corresponding to the relevant features. Moreover, the second part of Condition 5 ensures that the relevant features do not become overly dependent. Condition 6 indicates that the irrelevant features cannot have a strong effect on the relevant features.

\begin{theorem}\label{thm_pwr_logistic}
	Assume that Conditions 2--3 and 5--6 hold. With asymptotic probability one, $\frac{\left|\hat{\mathcal{S}} \cap \mathcal{S} \right|}{\left|\mathcal{S}\right|} \ge 1-\mathit{O}(b_n^{-1})$ for some $b_n \rightarrow \infty$, i.e. Power($\hat{\mathcal{S}}$) $\rightarrow 1$ as $n \rightarrow \infty$.
\end{theorem}

\begin{proof}
	By Theorem 1, Lemma 3 and the proof of Proposition 1 in \citet{ravikumar10}, if Conditions 5--6 are satisfied by the population Fisher information matrix $Q^*$, and $\lambda \ge \frac{16(2-\alpha)}{\alpha}\sqrt{\frac{\log p}{n}}$ for $\alpha$ in Condition 6, then 
	\begin{equation}
	||\hat{\beta}(\lambda)-\beta_T||_2 \leq \frac{5}{C_{\min}}\sqrt{s}\lambda
	\end{equation}
	with probability greater than $1-2exp\left\{-c\lambda^2 n \right\}$ for some positive constant $c$. Therefore, with high probability $||\hat{\beta}(\lambda)-\beta_T||_1=\mathcal{O}(s\lambda) $. Then following same arguments as in the proof of Theorem~\ref{thm_pwr_linear} with the use of Conditions 2--3, we have asymptotic power equal to one in this case as well. 
\end{proof}

\subsection{Approximation Construction}
Although the exact construction has desirable properties in terms of FDR control and asymptotic power, it has a major limitation with respect to computational cost: in Step 1 of exact construction, we need to calculate $2^p$ joint probabilities $\pi$ values, which is computational infeasible when $p$ is large. A computationally feasible version of construction method is in need for practical use. 

Inspired by the approximation construction in \citet{Candes17}, we modify the exact construction and propose a second-order approximation construction for Binary Knockoff procedure. Instead of ensuring the exchangeability condition \eqref{mxkf_swap} to hold exactly, we only ask for the first two moments in $\left(X, \tilde{X}\right)$ and $\left(X, \tilde{X}\right)_{\text{swap(S)}}$ to match. The payoff of violating the exact exchangeability condition to a small extent is a tremendous reduction in computational cost.

The second-order approximation construction of Binary Knockoffs is different from the exact construction in the first three steps. Here are the differences:
\begin{itemize}
	\item In \textbf{Step 1}, we choose a different combination of $\xi := (\eta, \gamma)$. For any index subset $I \subset \{1, \ldots, 2p\}$ where $|I|\leq 2$, we choose $\xi^{\left(X, \tilde{X}\right)_{I}}=\eta^{\left(X, \tilde{X}\right)_{I}}$, where $\left(X, \tilde{X}\right)_{I}$ corresponds to the superscript of $\eta$ defined in Section~\ref{2trans}. For any index subset $J \subset \{1, \ldots, 2p\}$ where $|J|> 2$, we choose $\xi^{\left(X, \tilde{X}\right)_{J}}=\gamma^{\left(X, \tilde{X}\right)_{J}}$, where $\left(X, \tilde{X}\right)_{J}$ corresponds to the superscript of $\gamma$ defined in Section~\ref{2trans}. In contrast to the Step 1 in the exact construction, the number of $\eta$ values in the mixed parameterization is only $2p \choose 2$ $+2p$.
	
	\item In \textbf{Step 2}, we calculate $\eta^{X_1}, \ldots, \eta^{X_p}, \eta^{X_1X_2}, \ldots, \eta^{X_{p-1}X_{p}}$, which consists of $\frac{p+p^2}{2}$ values compared to $2^p$ values in the exact construction.
	
	\item In \textbf{Step 3}, for any index subset $\mathcal{I} \subset \{1, \ldots, 2p\}$ where $|\mathcal{I}|\leq 2$ and for any swapping index subset $S \subset \{1, \ldots, p\}$, we set 
	\begin{equation} \label{exchangeable_2nd_order}
	\eta^{\left(X, \tilde{X}\right)_{\mathcal{I}}}=\eta^{\left\{\left(X, \tilde{X}\right)_{\text{swap(S)}}\right\}_{\mathcal{I}}}
	\end{equation}
	by using the calculated $\eta$ values in the modified Step 2.
	
	For the non-identifiable $\eta$'s, we set them to be some arbitrary values as long as \eqref{exchangeable_2nd_order} holds. And we still recommend setting all the $\gamma$ values to be zeros. 
\end{itemize}

The condition \eqref{exchangeable_2nd_order} ensures that the first two moments of $\left(X, \tilde{X}\right)$ and $\left(X, \tilde{X}\right)_{\text{swap(S)}}$ are matched. This construction is not exact because marginalizing the constructed joint distribution of $\left(X, \tilde{X}\right)$ over $\tilde{X}$ does not give back the given distribution of $X$. Therefore, the exchangeability condition \eqref{mxkf_swap} does not hold exactly. However, we show in the simulations that the approximation approach robustly controls FDR in practice.

\subsection{Parameters Unknown} \label{subsec:unknownIsing}

The ideal scenario considered in previous part may not be realistic all the time since the knowledge of the covariates distribution may not be available. Even though we model the covariate distribution using an Ising model, the true parameters are often unknown. Then it is natural to ask the question whether our Ising Knockoff procedure still controls FDR and holds the properties of the power if we use an estimated Ising parameter matrix for knockoff construction. Similar to \citet{fan17}, we may consider a data-split approach where half of the data are used for estimating $\Theta^*$ as $\hat{\Theta}$, and another half of the data for conducting Knockoff procedure. In practice, however, the data-split procedure may not be necessary as noted in the simulations of \citet{fan17} that FDR is still controlled without data-split. In our simulation we will also show that the FDR is controlled without the data-split step. 


Note that in the previous power analyses, only Condition 4 for linear regression and Condition 5--6 for logistic regression involve the augmented variable $X^*$ that contains the knockoffs. These three conditions are imposed directly on the expectations in terms of $X^*$. Therefore, if they hold for $X^*$ obtained from using the estimated Ising parameters, the power analysis conclusions will be the same. However, we need a further analysis of the FDR control when using $\hat{\Theta}$ for generating the knockoffs. 

\subsubsection{FDR Analysis}
Denote the FDR function of using the estimated parameter $\hat{\Theta}$ to be $\fdr(\hat{\Theta})$ and the FDR function of using the true parameter $\Theta^*$ to be $\fdr(\Theta^*)$. \citet{fan17} uses some Lipschitz function for analyzing the FDR control when using estimated precision matrix of a Gaussian graphical model, however, it is not easy to extend their idea in our case. We may mimic the way in \citet{fan17} by proposing a strong condition as the following:

\begin{condition} \label{lipschitz}
	There exists some constant $L>0$ such that for all $\left|\left|\hat{\Theta}-\Theta^*\right|\right|_{F}=\mathcal{O}(c_n)$ with $c_n \rightarrow \infty$, 
	\begin{equation}
	\left| \fdr(\hat{\Theta})-\fdr(\Theta^*) \right| \leq L \left|\left| \Theta^*-\hat{\Theta}  \right|\right|_{F}.
	\end{equation}
\end{condition}
By doing so, we bound the error term $\left| \fdr(\hat{\Theta})-\fdr(\Theta^*) \right|$ by the Frobenius norm of $\left|\left| \Theta^*-\hat{\Theta}  \right|\right|_{F}$. Note that a couple of existing methods are able to get an estimator of Ising parameters that satisfies the condition $\left|\left|\hat{\Theta}-\Theta^*\right|\right|_{F}=\mathcal{O}(c_n)$. For example, \citet{xue12} proposes an estimator of $\Theta^*$ that with probability tending to 1, $\left|\left| \hat{\Theta}-\Theta^*  \right|\right|_{F} = \mathcal{O}(\sqrt{\frac{s_1}{n}})$ for some constant $s_1$. Therefore, if Condition~\ref{lipschitz} holds, with high probability the estimated Ising Knockoff procedure can asymptotically control FDR at a target level. The difficulty remained is to check whether Condition~\ref{lipschitz} holds or not. We leave this part for future study.

\section{Empirical Results}
\subsection{Simulations}
\textbf{Simulation setup.} We compare the second-order approximation method of Binary Knockoff procedure with the approximation method of Model-X Knockoff procedure proposed by \citet{Candes17}. Both linear regression model and logistic regression model are considered in simulations. Note that in real data applications we usually do not know the true parameters of the features distribution. To show that our second-order approximation method has a robust performance on FDR control, in simulations we first estimate the first two moments of $X$ via sample mean and sample variance, and then use these estimated first two moments in the second-order approximation method. A similar estimation procedure is used in the real data analysis in next section. 

In a low dimensional linear regression model setup, we generate $n=400$ samples for 40 subgroups of features and each subgroup contains five features generated from an Ising model (i.e. $p=200$). We randomly set 30 out of 200 coefficients $\beta_j$'s to be $\pm L$ with $L$ ranging from $0.2$ to $0.5$, and all the rest coefficients are set to be zeros. In a high dimensional linear regression model setup, we generate $n=400$ samples for 120 subgroups of features and each subgroup contains five features generated from an Ising model (i.e. $p=600$). We randomly set 30 out of 600 coefficients $\beta_j$'s to be $\pm L$ with $L$ ranging from $0.2$ to $0.5$, and all the rest coefficients are set to be zeros. In a logistic regression model setup, we generate $n=400$ samples for 40 subgroups of features and each subgroup contains five features generated from an Ising model (i.e. $p=200$). We randomly set 30 coefficients $\beta_j$'s to be $\pm L$ with $L$ ranging from $0.5$ to $2.5$, and all the rest coefficients are zeros. In all three setting, the target FDR control level is $0.2$.

\begin{figure*} 
	\begin{center}
		\includegraphics[width=0.4\textwidth]{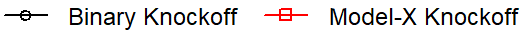}
	\end{center}
	\includegraphics[width=0.44\textwidth]{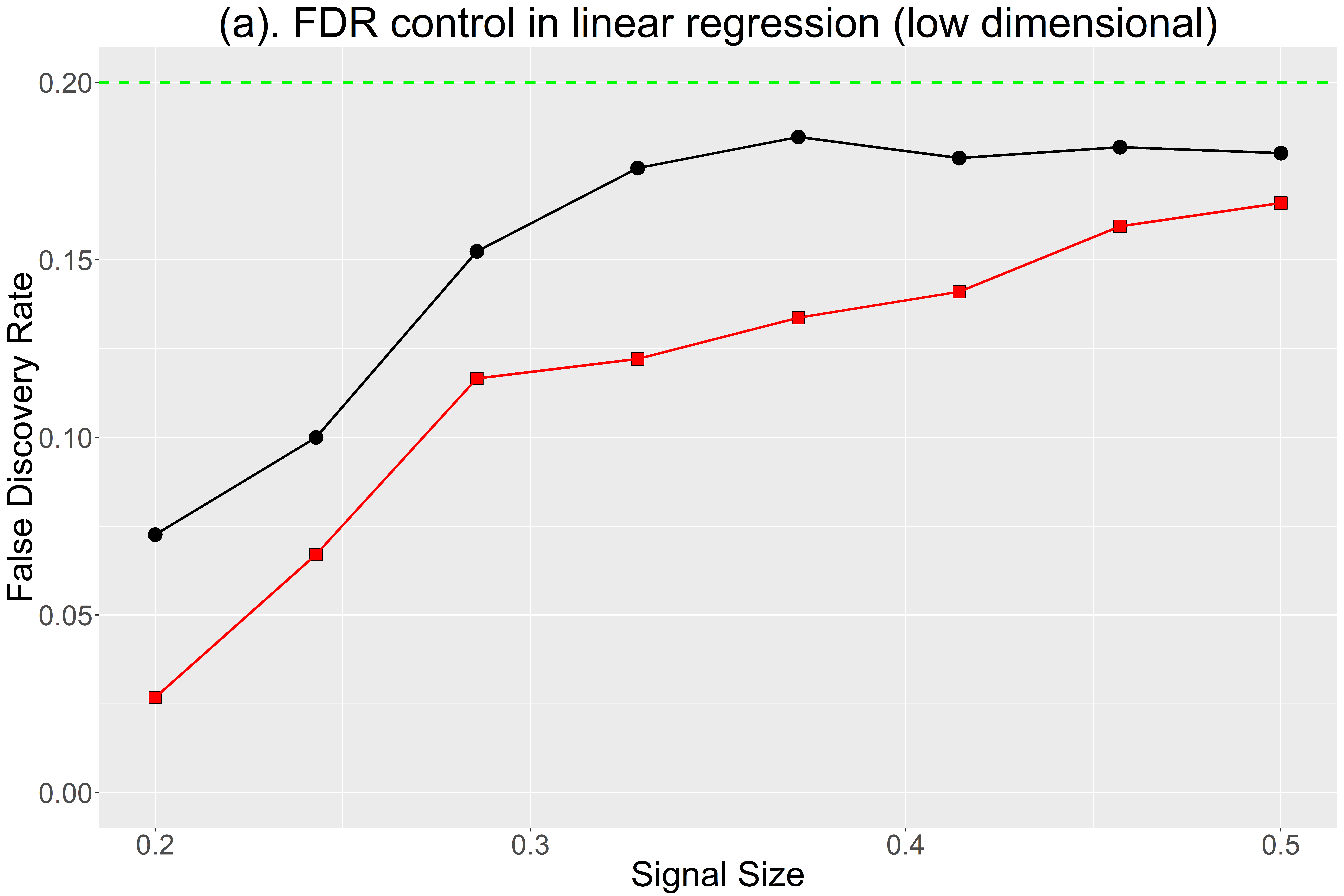} \includegraphics[width=0.44\textwidth]{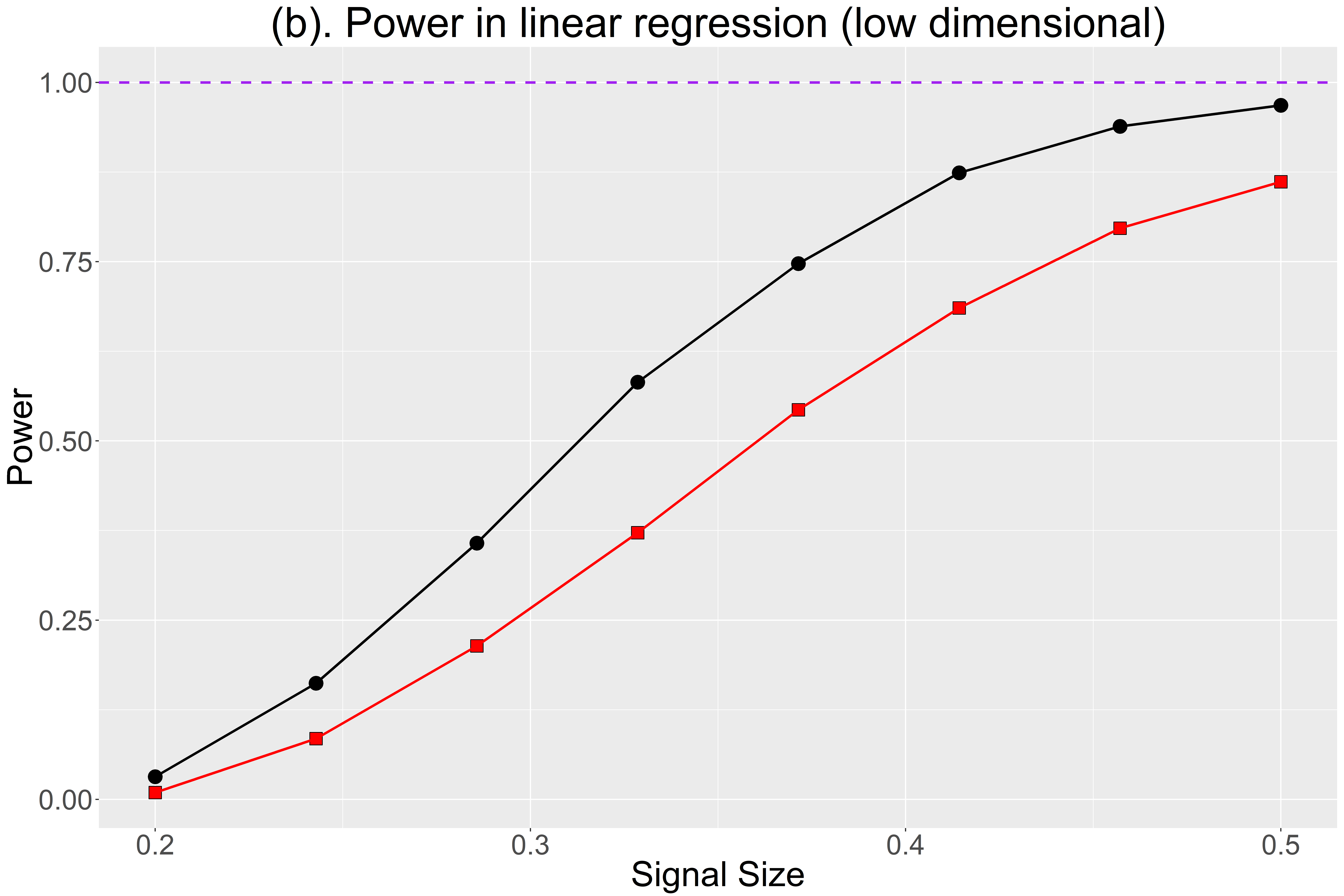}\hfill
	\includegraphics[width=0.44\textwidth]{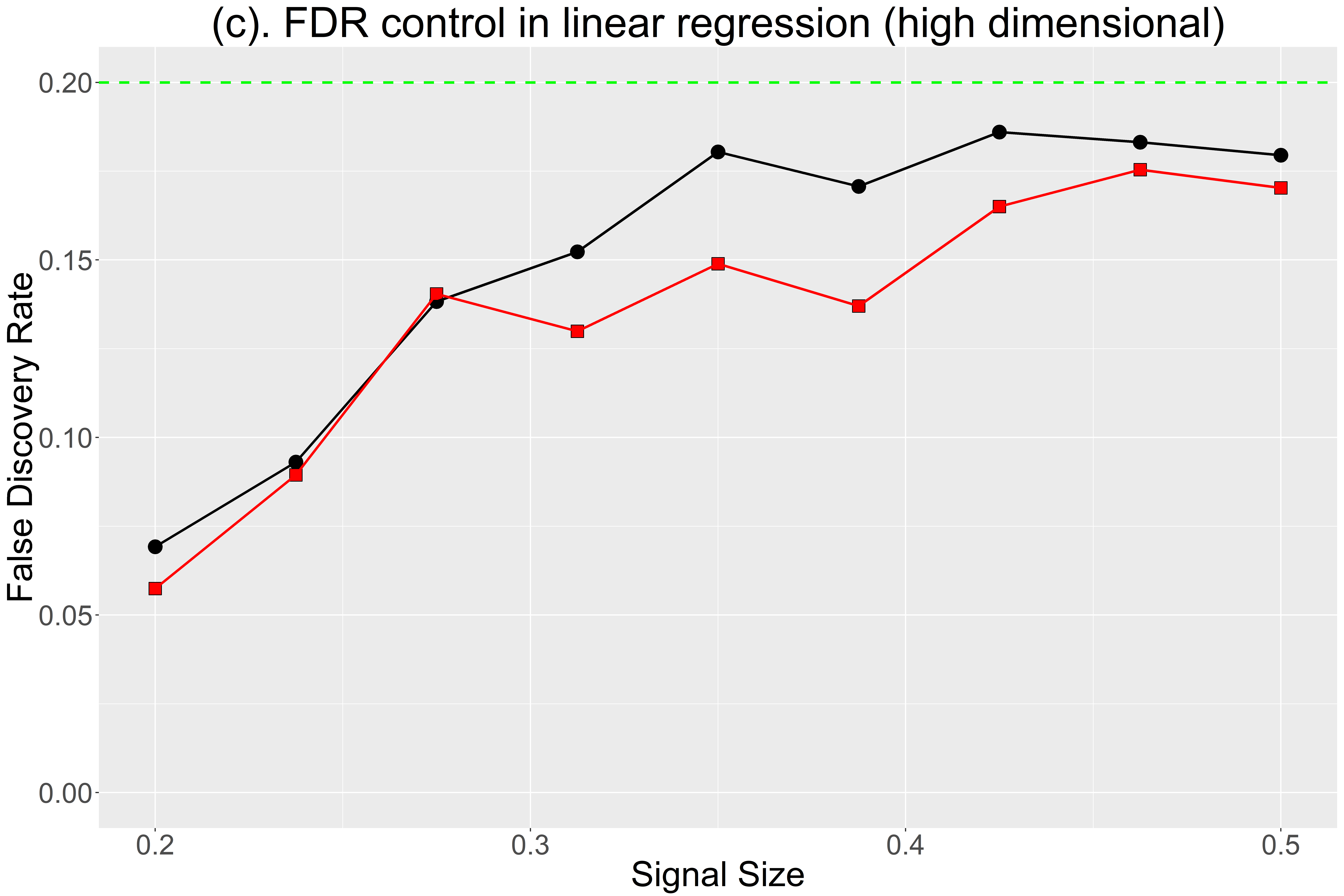} \includegraphics[width=0.44\textwidth]{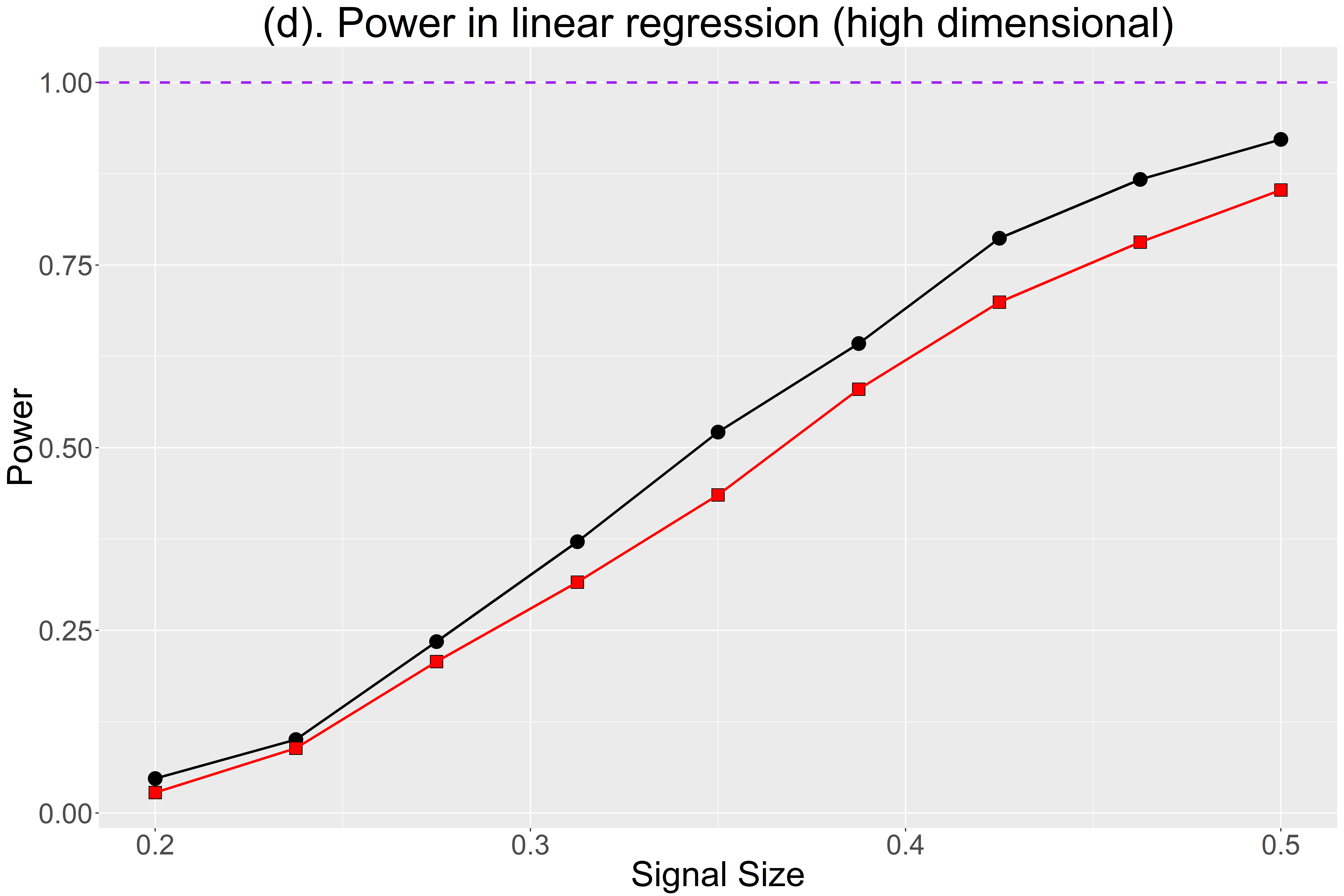}\hfill
	\includegraphics[width=0.44\textwidth]{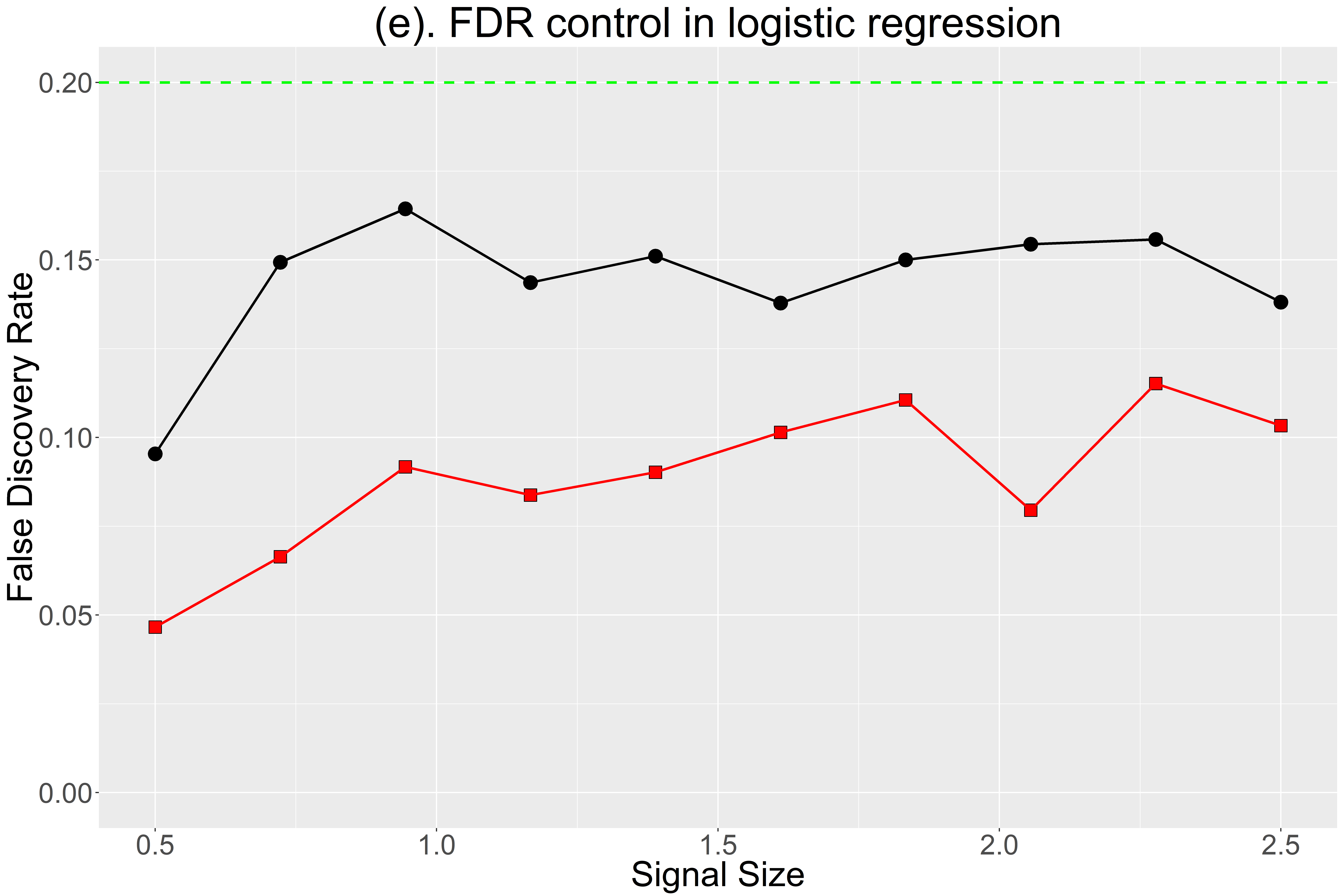} \includegraphics[width=0.44\textwidth]{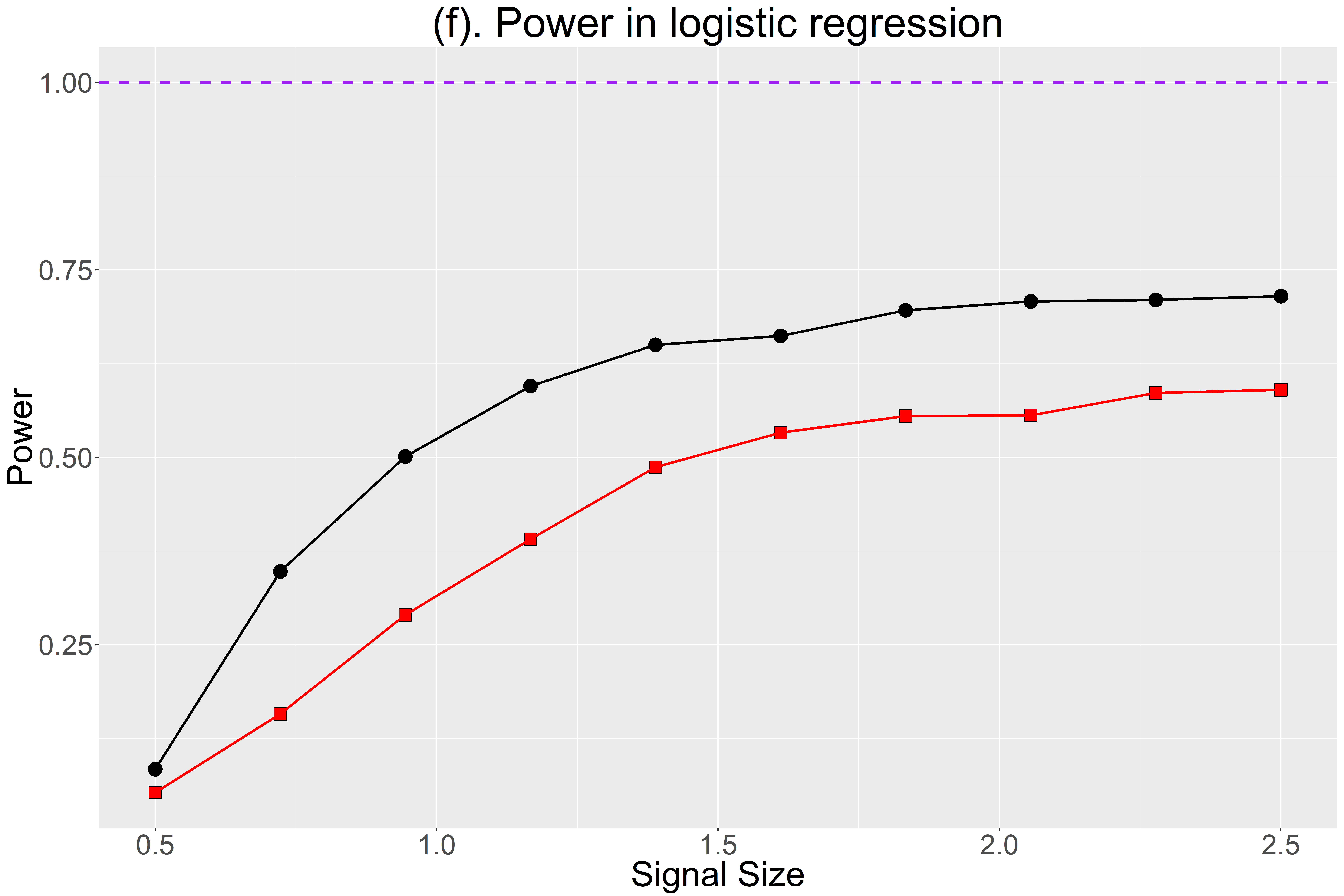}
	\caption{\label{sim} Simulation results from second-order approximation Binary Knockoff and approximation Model-X Knockoff. For ($a$) and ($b$), the model is linear regression with $n=400$ and $p=200$. For ($c$) and ($d$), the model is linear regression with $n=400$ and $p=600$. For ($e$) and ($f$), the model is logistic regression with $n=400$ and $p=200$.}
\end{figure*}

Although the second-order approximation construction violates the exchangeability condition \eqref{mxkf_swap}, we see from Figure~\ref{sim} that it still controls FDR under a pre-specified level in practice. Moreover, our approximation method of Binary Knockoff procedure has much higher power than the approximation method of Model-X Knockoff in all three simulation scenarios. One major difference between these two approximation methods is that the knockoffs generated by Binary Knockoff procedure are binary while the knockoffs constructed by Model-X Knockoff are continuous. It is more natural and reasonable to construct binary knockoffs for binary features; this may partially explain the gain of power in Binary Knockoff procedure in our simulations.

\subsection{Real Data}
We also illustrate the practical utility of Binary Knockoff procedure using a HIV antiretroviral therapy (ART) susceptibility dataset from the Stanford HIV drug resistance database. This dataset contains virus mutation information at protease residues for $702$ isolates from the plasma of HIV-1-infected patients. \citet{rhee06} and \citet{wu10} have used this dataset for studying the association between protease mutations and susceptibility to ART drugs. It has also been used in \citet{xue12} to study the graphical model of the protease residues, and \citet{xue12} model these protease residues using Ising graphical models.

We treat the protease residues as features and the amprenavir (APV) level as the response, and assume that their relationship follows a linear model. The mutations on each protease residue are recorded as binary values, so all the features of this dataset are binary, in which case the Binary Knockoff procedure is a more natural choice than other existing Knockoff procedures. Similar to a previous study in \citet{xue12}, we assume that all the features are generated from an Ising model, and can be partitioned into subgroups based on the stable edge graphs in the Figure 2 of \citet{xue12}. Our analysis uses $p=19$ of the residues that have at least 20\% of the values to be 1.

We apply the proposed second-order approximation Binary Knockoff method on this real data for controlling FDR at the level of 0.2. The first two moments of the features $X$ are estimated via sample mean and sample variance. We also provide the variable selection result by the approximation Model-X Knockoff procedure for comparison.

Table~\ref{tab} summarizes the results of two approximation methods. Since we do not know the ground truth, we search over genome science literature to find claims that support the association between the APV susceptibility and the residues frequently selected in this table. For example, \citet{mittal13} studies the association between APV and mutations at residue 50. The Table 1 in \citet{Martinez12} presents the APV resistance mutations at residue 33 and 36. Moreover, the Figure 1 in \citet{rhee06} shows their study about the association between APV and some residues listed in our Table~\ref{tab}. Most of the frequently selected residues in Table~\ref{tab} have literature supporting their association with APV, so we argue that most of the residues listed in Table~\ref{tab} are not false discoveries. In addition, both Knockoff procedures tend to select same residues, while the Binary Knockoff procedure has much higher selection frequencies than the existing Knockoff procedure. This result indicates that the Binary Knockoff procedure has a higher power, which matches the comparison results in the previous simulation studies. 

\begin{table}
	\centering
	\begin{tabular}{|p{1cm}|p{2.5cm}|p{2.5cm}|}
		\hline
		Residue & Selection by BKF &  Selection by MKF  \\
		\hline
		No.33 & 82\% &  70\%  \\
		\hline
		No.84 & 82\% &  70\%  \\
		\hline
		No.46 & 82\% &  68\%  \\
		\hline
		No.13 & 80\% &  63\%  \\
		\hline
		No.36 & 77\% & 64\%  \\
		\hline
		No.54 & 76\% & 63\%  \\
		\hline
		No.77 & 71\% & 52\%  \\
		\hline
		No.50 & 70\% & 52\%  \\
		\hline
	\end{tabular}
	\caption{\label{tab} The table of selection frequencies by Binary Knockoff procedure (BKF) and by Model-X Knockoff procedure (MKF) for residues being selected more than half times by both methods. }
\end{table}

\section{Conclusions and Discussions}
In this paper, we proposed Binary Knockoff procedure, an FDR controlled variable selection method tailored to binary features in regression framework. Since Ising model is commonly adopted for modeling the relationship among binary variables and has gained popularity in machine learning literature, this is a natural alternative to the Model-X knockoff in \citet{Candes17} and RANK in \citet{fan17} in the binary features setting. We provide both exact construction and second-order approximation construction of Binary Knockoff procedure. The exact construction leads to attractive theoretical results of FDR control and asymptotic power, and we show in empirical results that the second-order approximation method also controls FDR well in practice. 

We note that the way of constructing Binary Knockoffs in this paper can be easily extended to features generated from multivariate Bernoulli model, as the Ising model is a special case of the multivariate Bernoulli model. We expect that Ising model is probably useful enough for most practical applications. 

In spite of the good theoretical properties and empirical performance, our current proposal still have some limitations and thus can be improved in future research work. The inversion algorithm \citep{glonek96} we used in Step 4 of the construction procedure requires a good initial value for convergence. And it does not guarantee a valid output $\pi$ (i.e. all components of $\pi$ are non-negative) if there are some extreme $\eta$ values in input. It may happen when some binary features have very few $1$'s or $0$'s in a large sample, which indicates an extreme $\eta$ value is possible during the calculation in Step 2. To our best knowledge, this problem has not been solved in literature related to multivariate logistic regression model where the transformation $\pi \to \eta$ is frequently used. We leave this problem to future research work.

\bibliographystyle{apalike}
\bibliography{ref}

\begin{thebibliography}{}

\bibitem[Barber and Cand\`es, 2015]{barber15}
Barber, R.~F. and Cand\`es, E.~J. (2015).
\newblock Controlling the false discovery rate via knockoffs.
\newblock {\em Annals of Statistics}, 43(5):2055--2085.

\bibitem[B\"uhlmann and van~de Geer, 2011]{bv11}
B\"uhlmann, P. and van~de Geer, S. (2011).
\newblock Statistics for high-dimensional data: Methods, theory and
  applications.

\bibitem[Cand\`es et~al., 2017]{Candes17}
Cand\`es, E.~J., Fan, Y., Jason, L., and Lv, J. (2017).
\newblock Panning for gold: Model-x knockoffs for high-dimensional controlled
  variable selection.
\newblock {\em arXiv}.

\bibitem[Dai et~al., 2013]{Dai13}
Dai, B., Ding, S., and Wahba, G. (2013).
\newblock Multivariate bernoulli distribution.
\newblock {\em Bernoulli}, 19(4):1465--1483.

\bibitem[Fan and Li, 2001]{Fan01}
Fan, J. and Li, R. (2001).
\newblock Variable selection via nonconcave penalized likelihood and its oracle
  properties.
\newblock {\em Journal of the American Statistical Association},
  96(456):1348--1360.

\bibitem[Fan et~al., 2017]{fan17}
Fan, Y., Demirkaya, E., Li, G., and Lv, J. (2017).
\newblock Rank: Large-scale inference with graphical nonlinear knockoffs.
\newblock {\em arXiv}.

\bibitem[Fan et~al., 2016]{fan16}
Fan, Y., Kong, Y., Li, D., and Lv, J. (2016).
\newblock Interaction pursuit with feature screening and selection.
\newblock {\em arXiv}.

\bibitem[Gao et~al., 2018]{gao18}
Gao, C., Sun, H., Wang, T., Tang, M., Bohnen, N., M\"uller, M., Herman, T.,
  Giladi, N., Kalinin, A., Spino, C., Dauer, W., Hausdorff, J., and Dinov, I.
  (2018).
\newblock Model-based and model-free machine learning techniques for diagnostic
  prediction and classification of clinical outcomes in parkinson’s disease.
\newblock {\em Scientific Report}, (1):7129.

\bibitem[Glonek, 1996]{glonek96}
Glonek, G. (1996).
\newblock A class of regression models for multivariate categorical responses.
\newblock {\em Biometrika}, (83):15--28.

\bibitem[Ising, 1925]{Ising1925}
Ising, E. (1925).
\newblock Beitrag zur theorie des ferromagnetismus.
\newblock {\em Z. Physik}, 31(4):253--258.

\bibitem[Martinez-Cajas et~al., 2012]{Martinez12}
Martinez-Cajas, J., Wainberg, M., Oliveira, M., Asahchop, E., Doualla-Bell, F.,
  Lisovsky, I., Moisi, D., Mendelson, E., Grossman, Z., and Brenner, B. (2012).
\newblock The role of polymorphisms at position 89 in the hiv-1 protease gene
  in the development of drug resistance to hiv-1 protease inhibitors.
\newblock {\em Journal of Antimicrobial Chemotherapy}, 67(4):988--994.

\bibitem[McCullagh and Nelder, 1989]{mccullagh89}
McCullagh, P. and Nelder, J. (1989).
\newblock {\em Generalized Linear Models}.

\bibitem[Mittal et~al., 2013]{mittal13}
Mittal, S., Bandaranayake, R., King, N., Prabu-Jeyabalan, M., Nalam, M.,
  Nalivaika, E., Yilmaz, N., and Schiffer, C. (2013).
\newblock Structural and thermodynamic basis of amprenavir/darunavir and
  atazanavir resistance in hiv-1 protease with mutations at residue 50.
\newblock {\em Journal of Virology}, 87(8):4176--4184.

\bibitem[Ravikumar et~al., 2010]{ravikumar10}
Ravikumar, P., Wainwright, M.~J., and Lafferty, J.~D. (2010).
\newblock High-dimensional ising model selection using l1-regularized logistic
  regression.
\newblock {\em Annals of Statistics}, 38(3):1287--1319.

\bibitem[Rhee et~al., 2006]{rhee06}
Rhee, S.-Y., Taylor, J., Wadhera, G., Ben-Hur, A., Brutlag, D., and Shafer, R.
  (2006).
\newblock Genotypic predictors of human immunodeficiency virus type 1 drug
  resistance.
\newblock {\em Proceedings of the National Academy of Sciences of the United
  States of America}, 103(46):17355--17360.

\bibitem[Sesia and Cand\`es, 2018]{Sesia18}
Sesia, M. and Cand\`es, E.~J. (2018).
\newblock Gene hunting with hidden markov model knockoffs.
\newblock {\em Biometrika}.

\bibitem[Shao, 1993]{Shao93}
Shao, J. (1993).
\newblock Linear model selection by cross-validation.
\newblock {\em Journal of the American Statistical Association},
  88(422):486--494.

\bibitem[Tibshirani, 1996]{Tibshirani96}
Tibshirani, R. (1996).
\newblock Regression shrinkage and selection via the lasso.
\newblock {\em Journal of the Royal Statistical Society. Series B},
  58(1):267--288.

\bibitem[Weinstein et~al., 2018]{Weinstein18}
Weinstein, A., Barber, R., and Cand\`es, E.~J. (2018).
\newblock A power and prediction analysis for knockoffs with lasso statistics.
\newblock {\em arXiv}.

\bibitem[Wu et~al., 2010]{wu10}
Wu, M., Cai, T., and Lin, X. (2010).
\newblock Testing for regression coefficients in lasso regularized regression.
\newblock {\em Technical report, Harvard University}.

\bibitem[Xiao et~al., 2017]{xiao17}
Xiao, Y., Angulo, T., Friedman, J., Waldor, M., Weiss, S., and Liu, Y. (2017).
\newblock Mapping the ecological networks of microbial communities from
  steady-state data.
\newblock {\em bioRxiv}, page 150649.

\bibitem[Xie et~al., 2018]{xie18}
Xie, Y., Chen, N., and Shi, X. (2018).
\newblock False discovery rate controlled heterogeneous treatment effect
  detection for online controlled experiments.
\newblock {\em Proceedings of the 24th ACM SIGKDD International Conference on
  Knowledge Discovery \& Data Mining}, pages 876--885.

\bibitem[Xue et~al., 2012]{xue12}
Xue, L., Zou, H., and Cai, T. (2012).
\newblock Nonconcave penalized composite conditional likelihood estimation of
  sparse ising models.
\newblock {\em Annals of Statistics}, 40(3):1403--1429.

\bibitem[Yu and Feng, 2014]{Yu14}
Yu, Y. and Feng, Y. (2014).
\newblock Model selection via multifold cross validation.
\newblock {\em Journal of Computational and Graphical Statistics},
  23(4):1009--1027.

\bibitem[Zhang, 1993]{Zhang93}
Zhang, P. (1993).
\newblock Model selection via multifold cross validation.
\newblock {\em The Annals of Statistics}, 21(1):299--313.

\bibitem[Zou and Hastie, 2005]{Zou05}
Zou, H. and Hastie, T. (2005).
\newblock Regularization and variable selection via the elastic net.
\newblock {\em Journal of the Royal Statistical Society. Series B},
  67(2):301--320.

\end{thebibliography}
\end{document}